\documentclass{jaa}
\usepackage{graphicx}

\begin{document}\sloppy

%%paper title
%%For line breaks \\ can be used within title 
\title{[Rb/Zr] ratio in Ba stars as diagnostics of the companion AGB stellar mass}

%%author names are separated by comma (,) 
%%use \and before the last author name 
%%use a * along with the number separated by comma
%% for the  author for correspondence
%%\textsuperscript{number} is used for affiliation
%%\affilOne, \affilTwo etc., upto \affilTwentyfive is possible
%%Please note the first letter after \affil is capitalised in the command
%%

\author{J. Shejeelammal\textsuperscript{1,*} \and Aruna Goswami\textsuperscript{1,*}}
\affilOne{\textsuperscript{1}Indian Institute of Astrophysics, Bangalore 560034, India.\\}
%\affilTwo{\textsuperscript{2}Department of Q, University Z, Place Pincode, Country.}

%%escape two column mode for title, affiliation and abstract
%%by giving \twocolumn command as shown

\twocolumn[{

\maketitle

%%include \corres to print the corresponding author Email id
\corres{shejeelammal.j@iiap.res.in, aruna@iiap.res.in}

%%include \msinfo for
%%manuscript information such as
%%received, revised and accepted dates
%%
\msinfo{--}{--}

%%abstract
\begin{abstract}
Understanding the nucleosynthesis and evolution of Asymptotic Giant Branch (AGB) stars is of primary importance as they are the main producers of some of the key elements in the Universe. They are the predominant sites for the slow neutron-capture nucleosynthesis. The exact physical conditions and nucleosynthetic processes occuring at the interior of AGB stars are not clearly understood, that hinders a better understanding of the contribution of these stars to the Galactic chemical enrichment. The extrinsic stars, which are known to have received products of AGB phase of evolution via binary mass transfer mechanisms, form vital tools to trace the AGB nucleosynthesis.  The [Rb/Zr] ratio is an important diagnostic to understand the average neutron density at the s-process site and  provide important clues to the mass of the companion AGB stars. In this work we have  presented the estimate of [Rb/Zr] ratios, based on the high resolution spectroscopic analysis for a sample of Ba stars, and discussed  how it can be used to understand the characteristics of the AGB star. Results from an analysis based on parametric model to confirm the mass of the companion AGB star are also presented. 
\end{abstract}

%%insert keywords separated by 3 hyphens using \keywords{words}

}]
%%close the twocolumn escape here

%%include \doinum{number}for the DOI number in the header
%%include \volnum{number} for the volume number in the header
%%include \year{yyyy} for  year of publication in the header
%%include \pgrange{num--num} page range of article in the header
%%include \artcitid{num} for the article citation id
%%include \lp to print last page of the article
%%include \setcounter{page}{pagenum} for the exact starting page of the article

\doinum{12.3456/s78910-011-012-3}
\artcitid{\#\#\#\#}
\volnum{000}
\year{0000}
\pgrange{1--}
\setcounter{page}{1}
\lp{1}

\section{Introduction}
Barium (Ba II) stars are peculiar G and K type stars which were first identified  
by Bidelman and Keenan (1951). 
Their surface chemical composition is characterized  by  overabundance of elements heavier than iron and C/O $<$ 1 (Barbuy et al. 1992, Allen and Barbuy 2006a, Drake and Pereira 2008, Pereira and Drake 2009). They exhibit abnormally strong lines of s-process (slow neutron-capture process)
elements such as Ba II at $\lambda$ 4554 \AA, Sr II at $\lambda$ 4077 \AA,
as well as enhanced CH, CN and $C_{2}$ molecular bands. They are mostly in their Main-Sequence and giant phase of stellar evolution.
Nucleosynthesis theories do not support occurrence of heavy element nucleosynthesis during the stellar evolutionary phases to which 
these stars belong.   

It is known that the Asymptotic Giant Branch (AGB) stars are the major producers of
s-process elements in the Universe (Busso et al. 1999). The s-process enriched materials produced in the 
interiors of the AGB stars are brought to the surface through Third Dredge-Up (TDU). 
The observed over abundance of heavy elements on the surface of the Ba stars could not be attributed to an 
intrinsic origin as they are not luminous enough to undergo s-process nucleosynthesis.  This posed a challenge to the existing nucleosynthesis theories.

The radial velocity monitoring studies of the Ba stars have shown that  85\% of the Ba stars are in binaries
(McClure et al. 1980, McClure 1983, 1984, McClure \& Woodsworth 1990, Udry et al. 1998a,b, Lucatello et al. 2005) 
with a now invisible white dwarf companion. Later studies confirmed that all the giant Ba stars 
are binaries (Jorissen et al 2019). A generally accepted scenario that explains the observed high abundances of
neutron-capture elements is a binary mass transfer picture. These stars are believed to have received via
binary mass transfer mechanisms the products of the companion stars produced during their AGB
phase of evolution. 
Hence, the chemical composition of this class of objects can be used to trace the
AGB nucleosynthesis at their corresponding metallicity.

The detailed chemical composition studies of 
the AGB stars can help  better understanding of the evolution of heavy elements in the Galaxy. 
The molecular contribution dominant spectra of the AGB stars make the derivation of 
exact elemental abundance difficult. In this regard, the spectra of the comparatively 
warmer, extrinsically s-process enhanced, Ba stars could make the derivation of elemental abundance much easier 
and  help  probing the s-process enrichment of the Galaxy. 

Several aspects of the Ba stars such as mass, abundance peculiarities, kinematics etc. have been extensively studied by many authors
(Allen \& Barbuy 2006a, Smiljanic et al. 2007, de Castro et al. 2016,  Yang et al. 2016, Mahanta et al. 2016, Karinkuzhi et al. 2018a,
Purandardas et al. 2019, Shejeelammal et al. 2020 and references therein). In addition to the probing of s-process enrichment of the Galaxy,
the surface chemical composition analysis of the Ba stars could also be used to characterize the initial mass of the 
companion AGB stars. 
%However, there is only a few such dedicated studies available in literature 
%(Shejeelammal et al. 2020, Karinkuzhi at al. 2018b, de Castro et al. 2016). 
Mass is one of the  important basic  parameters of stars; the neutron source and nucleosynthesis product distribution varies with
initial stellar mass.  
In case of AGB stars, there are two important neutron sources for the s-process in the He intershell: 
$^{13}$C($\alpha$, n)$^{16}$O reaction during the radiative inter-pulse period and
$^{22}$Ne($\alpha$, n)$^{25}$Mg reaction during the convective thermal pulses. 
$^{13}$C($\alpha$, n)$^{16}$O reaction is the dominant neutron source in
low-mass AGB stars with initial mass $\leq$ 3 M$_{\odot}$. 
The temperature required for the operation of this reaction is 
T $\geq$ 90 $\times$ 10$^{6}$ K and provides a neutron density 
N$_{n}$ $\sim$ 10$^{8}$ cm$^{-3}$ in a timescale of $\geq$ 10$^{3}$ years
(Straniero et al. 1995, Gallino et al. 1998, Goriely \& Mowlavi 2000, Busso et al. 2001).
The temperature required for the activation of $^{22}$Ne source is 
300$\times$10$^{6}$ K, which is achieved during the TPs in 
intermediate mass AGB stars (initial mass $\geq$ 4 M$_{\odot}$). 
It produces a neutron density N$_{n}$ $\sim$ 10$^{13}$ cm$^{-3}$ 
in a timescale of $\sim$ 10 years. The temperature required for 
the $^{22}$Ne source is reached in low-mass stars during the 
last few TPs providing N$_{n}$ $\sim$ 10$^{10}$ - 10$^{11}$ cm$^{-3}$
(Iben 1975, Busso et al. 2001).

Rb plays a unique role as diagnostics of the neutron density at the s-process site.
Rb is the only low neutron density branch available to the stellar spectroscopists as a neutron densitometer (Tomkin\& Lambert 1999). 
 Rb is produced only when the N$_{n}$ $>$ 5$\times$ 10$^{8}$ cm$^{-3}$,
otherwise Sr, Y, Zr etc. are produced. Hence, [Rb/Zr] ratio can be used as an indicator of mass of AGB stars. 
Theoretical models predict a negative value for [Rb/Zr] ratio in low-mass AGB stars and a positive 
value in intermediate-mass AGB stars (Abia et al. 2001, van Raai et al. 2012, Karakas et al. 2012). 
We have used the neutron density dependent [Rb/Zr] ratio in 
a sample of barium stars to infer the mass of companion AGB stars. Such studies in literature are scanty. 

In this work, we present results obtained from a detailed chemical composition analysis of 
four Ba stars: HD~32712, HD~36650, HD~179832 and HD~211173, based on high-resolution, high quality spectra. 
We have determined the abundances of heavy elements such as Rb, Sr, Y, Zr, Ba, La, Ce, Pr, Nd, Sm and Eu in these stars.
The details of the spectra, abundance analysis and the interpretations of the results are discussed in the subsequent sections. 

%\subsection{Subsection heading}
%Subsection text here. 

%\subsubsection{Subsubsection heading.} Subsubsection text goes here (Radhakrishnan {\em et al.} 1980).

\section{Data acquisition and data reduction}
\begin{figure}
\centering
\includegraphics[width=\columnwidth]{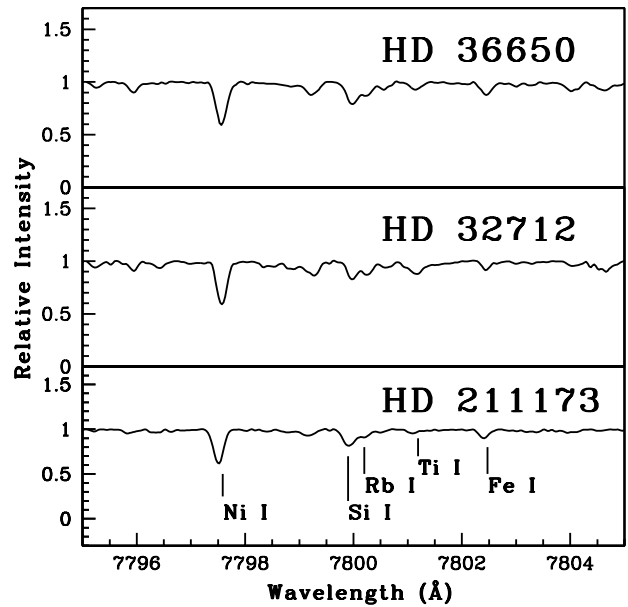}
\caption{ Sample spectra of the program stars in the  wavelength region 
7795 to 7805 {\bf  {\rm \AA}}.}\label{sample_spectra}
\end{figure}

The objects analysed in this study are taken from the barium star catalog of L\"u (1991).
The high resolution  ($\lambda/\delta\lambda \sim 48,000 $) 
FEROS (Fiber-fed Extended Range Optical Spectrograph attached to  the
1.52 m telescope of ESO at La Silla, Chile) spectra are used for  all the 
four objects.  The wavelength coverage  spans from 3520 - 9200 {\rm \AA}.
The data reduction is performed using the 
basic tasks in Image Reduction and Analysis Facility (IRAF) software.
A few sample spectra are shown in Figure \ref{sample_spectra}.

\section{Data analysis}
\subsection{Determination of stellar atmospheric parameters}
The data analysis is performed using the most recent version of the
radiative transfer code MOOG (Sneden 1973) based on the 
assumptions of Local Thermodynamic Equilibrium (LTE).
The stellar atmospheric parameters, the effective temperature T$_{eff}$, the surface gravity log g, 
micro-turbulent velocity $\zeta$, and the metallicity [Fe/H] are determined using a set of clean, unblended 
Fe I and Fe II lines with excitation potential in 
the range 0.0 - 6.0 eV and equivalent width 20 - 180 m\AA. An initial 
model atmosphere is selected from the Kurucz grid of model atmosphere 
with no convective overshooting (http://cfaku5.cfa.hardvard.edu/) using 
the photometric temperature estimate and the initial guess of log g value for 
giants/dwarfs. The final model atmosphere is obtained through an iterative
method from the initially selected one. The effective temperature is determined 
by forcing the slope of abundances versus 
the excitation potential of the measured Fe I lines to zero. 
The micro-turbulent velocity at that particular temperature is fixed to be that value
for which there is  no dependence of the abundances derived from the Fe I lines on the
reduced equivalent width. The surface gravity is obtained by demanding the abundances 
derived from both Fe I and Fe II lines give nearly same values at the selected effective 
temperature and microturbulent velocity. The abundances obtained from the Fe I and Fe II lines
give the metallicity. With this finally adopted model atmosphere, the further abundance analysis 
is carried out. 
  
\subsection{Abundances of heavy elements}
\begin{figure}
\centering
\includegraphics[width=\columnwidth]{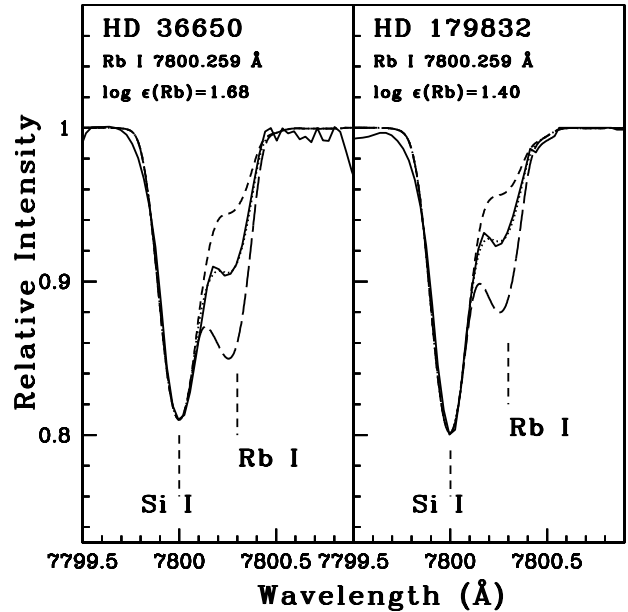}
\caption{ Synthesis of Rb I line around 7800 {\rm \AA}. Dotted line 
represents synthesized spectra and the solid line indicates the 
observed spectra. Short dashed line represents the synthetic spectra 
corresponding to $\Delta$[Rb/Fe] = $-$0.3 and long dashed line is 
corresponding to $\Delta$[Rb/Fe] = +0.3} \label{Rb_7800}
\end{figure}

\begin{figure}
\centering
\includegraphics[width=\columnwidth]{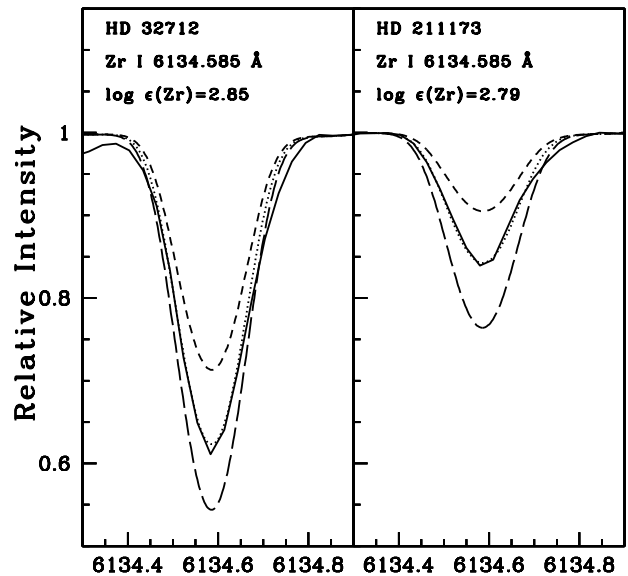}
\caption{ Synthesis of Zr I line at 6134.585 {\rm \AA}. Dotted line 
represents synthesized spectra and the solid line indicates the observed 
spectra. Short dashed line represents the synthetic spectra 
corresponding to $\Delta$[Zr/Fe] = $-$0.3 and long dashed line is 
corresponding to $\Delta$[Zr/Fe] = +0.3} \label{Zr_synth}
\end{figure}

The abundance of Rb is derived  using the spectral synthesis 
calculation  of Rb I resonance line at 7800.259 \r{A} (Figure \ref{Rb_7800}).
We could not detect the Rb I lines in the warmer program stars. 
The Rb I resonance line at 7947.597 \r{A} is not usable for the abundance estimation.
The hyper-fine components of Rb is taken from Lambert \& Luck (1976).
The Sr abundance is derived from the spectral synthesis calculation of
Sr I line at 4607.327 \r{A}. The Y I line at 6435.004 \r{A} is used to derive
the Y I abundances while Y II abundances are derived from the measured equivalent width of
several Y II lines. We have derived the Zr abundance from the spectral synthesis of 
Zr I line at 6134.585 \r{A} (Figure \ref{Zr_synth})
and from the equivalent width measurement of  several Zr II lines.
The abundances of Ba and La are derived from the spectral synthesis calculation of 
Ba II line at 5853.668 \r{A} and  La II line at 4921.776 \r{A}. 
The hyper-fine components of Ba are taken from Mcwilliam (1998) and La from Jonsell et al. (2006). 
To derive the abundances of elements Ce, Pr, Nd, Sm, equivalent width measurement of several 
singly ionized lines are used. The Eu abundances are derived from the spectral synthesis 
calculation of Eu II line at 6645.064 \r{A} by considering the hyper-fine components from 
Worely et al. (2013). 
 
\section{Results and Discussion}
\subsection{[Rb/Zr] ratio as a diagnostic of the neutron density at the s-process site}
\begin{figure}
\centering
\includegraphics[scale=0.15]{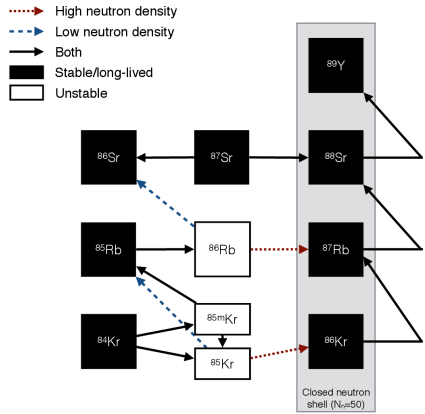}
\caption{s-process path from Kr to Zr. Stable isotopes are
represented by white box, while the black box represents 
the unstable ones. The solid arrows indicate the s-process path branches
specific to low neutron densities. The dashed arrows indicate the
s-process path branches followed under high neutron densities. The
open rectangular box surrounds the isotopes with magic neutron number N = 50 (van Raai et al. 2012).} 
\label{Kr_branching}
\end{figure}

Rb is an s-process element produced from the neutron capture of $^{84}$Kr. 
The s-process path going through Rb is shown in Figure \ref{Kr_branching}.
The  branching points at the 
unstable nuclei $^{85}$Kr and $^{86}$Rb controls the 
Rb production along the s-process nucleosynthesis path trough Rb. 
The probability of these unstable nuclei to 
capture the neutron before $\beta$-decaying determines the amount of Rb produced, 
which in turn depends on the neutron density at the s-process 
site (Beer \& Macklin 1989, Tomkin \& Lambert 1983, Lambert et al. 1995). 
When $^{84}$Kr undergoes neutron capture,  50\% of the flux goes to 
the ground state of $^{85}$Kr and other 50\% goes to the 
metastable state of $^{85}$Kr. Out of this metastable $^{85}$Kr, 
80\% decays to  $^{85}$Rb, while the remaining 20\% decays to 
its ground state. In effect,  40\% of the $^{84}$Kr+n produces $^{85}$Rb, 
where as the remaining 60\% results in the  production of ground 
state $^{85}$Kr. At higher neutron densities, N$_{n}$ $>$ 5$\times$10$^{8}$ 
n/cm$^{3}$, the long-lived $^{85}$Kr (t$_{1/2}$ $\sim$ 10.75 yrs) 
undergo neutron capture allowing the reaction 
$^{85}$Kr(n, $\gamma$)$^{86}$Kr(n, $\gamma$)$^{87}$Kr.
The short-lived $^{87}$Kr (t$_{1/2}$ $\sim$ 76.3 min) decays 
quickly to the  stable $^{87}$Rb. The other unstable isotope 
$^{86}$Rb (t$_{1/2}$ $\sim$ 18.63 days), which is produced by 
the neutron capture of $^{85}$Rb, directly produces $^{87}$Rb 
provided the neutron density is $\geq$10$^{10}$ n/cm$^{3}$.  
At lower neutron densities, $^{86}$Rb decays to $^{86}$Sr allowing for
$^{86}$Sr(n, $\gamma$)$^{87}$Sr(n, $\gamma$)$^{88}$Sr (Beer 1991, 
Lugaro \& Chieffi 2011).

The magic number of neutrons of $^{87}$Rb makes it stable against neutron capture. 
The smaller neutron-capture cross-section of $^{87}$Rb ($\sigma$ $\sim$ 15.7 mbarn at 30 KeV) 
compared to that of $^{85}$Rb ($\sigma$ $\sim$ 234 mbarn) (Heil 
et al. 2008a) favours the accumulation of $^{87}$Rb once it is formed. 
Therefore, the isotopic ratio $^{87}$Rb/$^{85}$Rb
could be  a direct indicator of the neutron density at 
the s-process site, which in turn  help  to infer the mass 
of the AGB star. But, it is impossible to distinguish the lines 
due to these two isotopes of Rb in the stellar spectra 
(Lambert \& Luck 1976, Garc\'ia-Hern\'andez et al. 2006).
However, the abundance of Rb relative to other elements in 
this region of the s-process path, such as Sr, Y, and Zr, 
can be used to estimate the average neutron density of the s-process.
Both the theoretical models and the observations have shown that the 
[Rb/Zr] has a negative value in AGB stars with M $<$ 3M$_{\odot}$ and
a positive value in massive AGB stars with mass M $>$ 3M$_{\odot}$ (Karakas et al. 2012, Plez et al. 1993, Lambert et al. 1995, 
Abia et al. 2001,  Garc\'ia-Hern\'andez et al. 2006, 2007, 2009, 
van Raai et al. 2012). 

\begin{figure}
\centering
\includegraphics[scale=0.28]{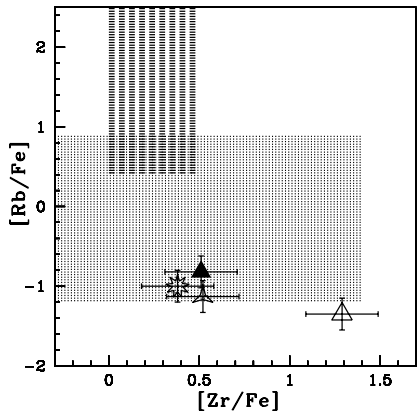}
\caption{The observed abundances [Rb/Fe] vs [Zr/Fe].
 HD~32712 (starred triangle), HD~36650 (filled triangle), 
HD~179832 (open triangle), and HD~211173 (nine-sided star). 
The region shaded with short-dashed line  and dots 
corresponds to the observed range of Zr and Rb in 
intermediate-mass  and low-mass AGB stars respectively 
in the Galaxy and the Magellanic Clouds 
(van Raai et al. 2012).}
\label{Rb_Zr}
\end{figure}

We could derive the [Rb/Zr] ratio in all the  four Ba stars, the estimated values are given in Table \ref{rb_zr}. 
The observed [Rb/Fe] and [Zr/Fe] ratios are  shown 
in Figure \ref{Rb_Zr}. The observed ranges of Rb and Zr in 
low- and intermediate-mass AGB stars (shaded regions) 
in the Galaxy and Magellanic Clouds are also shown for a 
comparison. The comparison shows clear evidence of consistency between 
observed abundances of Rb and Zr in the Ba stars and their counterparts 
normally observed in the 
low-mass AGB stars. The negative values of [Rb/Zr] ratio obtained in the 
program stars confirm the
low-mass companion for these stars. 
{\footnotesize
 \begin{table}
\caption{Estimates of [Rb/Fe], [Zr/Fe] \& [Rb/Zr]} \label{rb_zr}
\resizebox{\columnwidth}{!}{\begin{tabular}{|l|l|l|l|l|}
\hline\hline 
            &                      &            &            &               \\
Star name   & [Fe/H]               & [Rb/Fe]    & [Zr/Fe]    & [Rb/Zr]      \\ 
            &                      &($\pm$ 0.20 dex) & ($\pm$ 0.20 dex) & ($\pm$ 0.20 dex)                          \\
\hline
HD 32712    & $-$0.25$\pm$0.12     & $-$1.13    & 0.52       & $-$1.65      \\
HD 36650    & $-$0.02$\pm$0.12     & $-$0.82    & 0.51       & $-$1.33      \\
HD 179832   & +0.23$\pm$0.04       & $-$1.35    & 1.29       & $-$2.64       \\
HD 211173   & $-$0.17$\pm$0.10     & $-$1.00    & 0.38       & $-$1.38   \\
\hline
\end{tabular}} 
 \end{table}
 }

\subsection{Comparison with FRUITY models and a parametric model based 
analysis}
The observed abundance in the atmosphere of the 
Ba stars might not be the actual signature of the 
progenitor AGB stars. The accreted s-rich material
is mixed and diluted in the envelope of these 
secondary Ba stars. The diluted theoretical abundance on the 
surface of Ba stars is given as (Husti et al. 2009);

[X/Fe] = log(10$^{[X/Fe]^{ini}}$ . f + 10$^{[X/Fe]^{AGB}}$ . 10$^{-d}$) \\

 where d is the dilution factor, f = 1 - 10$^{-d}$ and [X/Fe]$^{AGB}$ is 
 the abundance of element X in the AGB. In this formulation it is assumed that
 both the AGB and the Ba stars are formed from the same cloud of
 interstellar medium (Husti et al. 2009).

\begin{figure}
\centering
\includegraphics[scale=0.28]{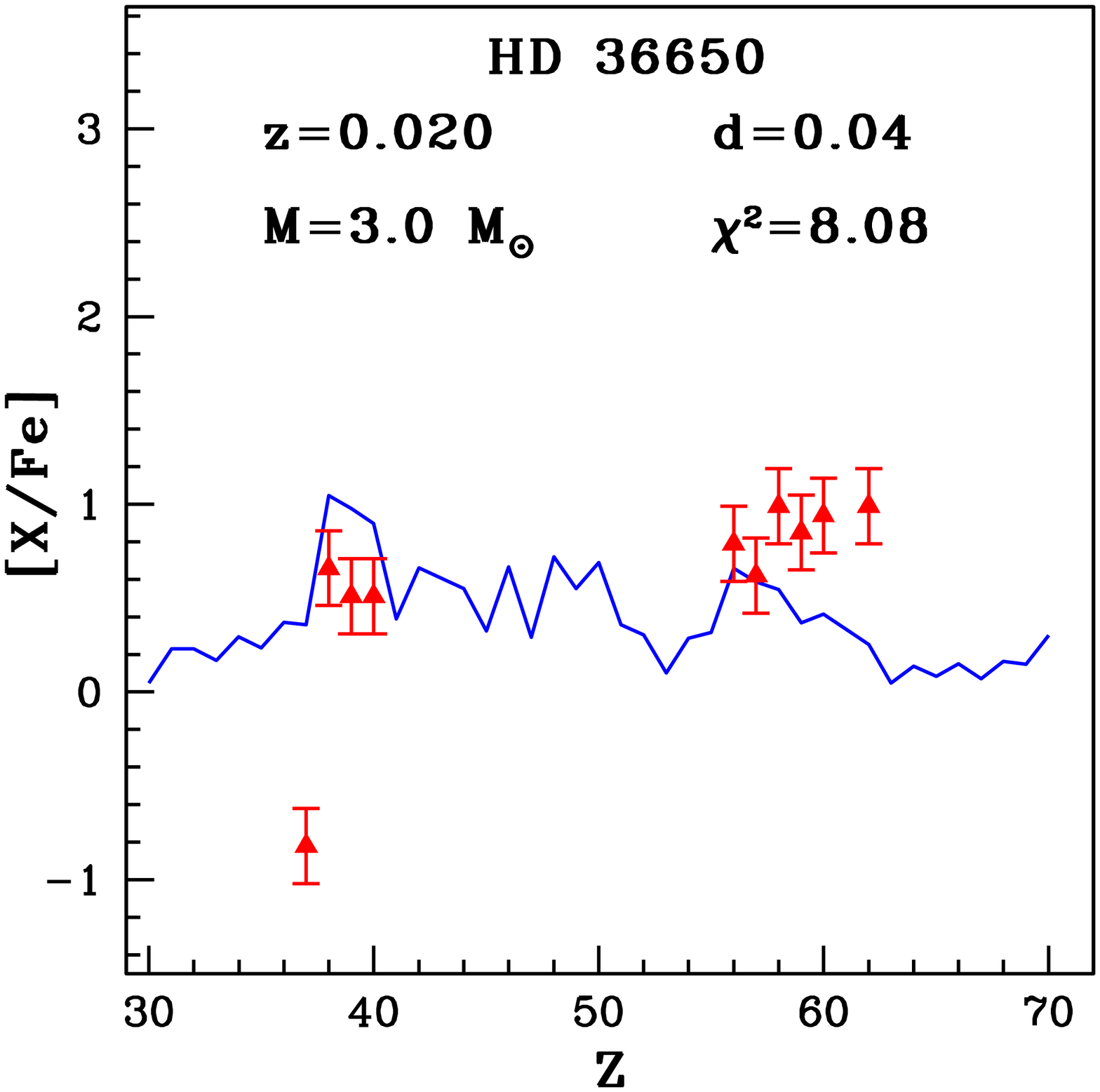}
\includegraphics[scale=0.28]{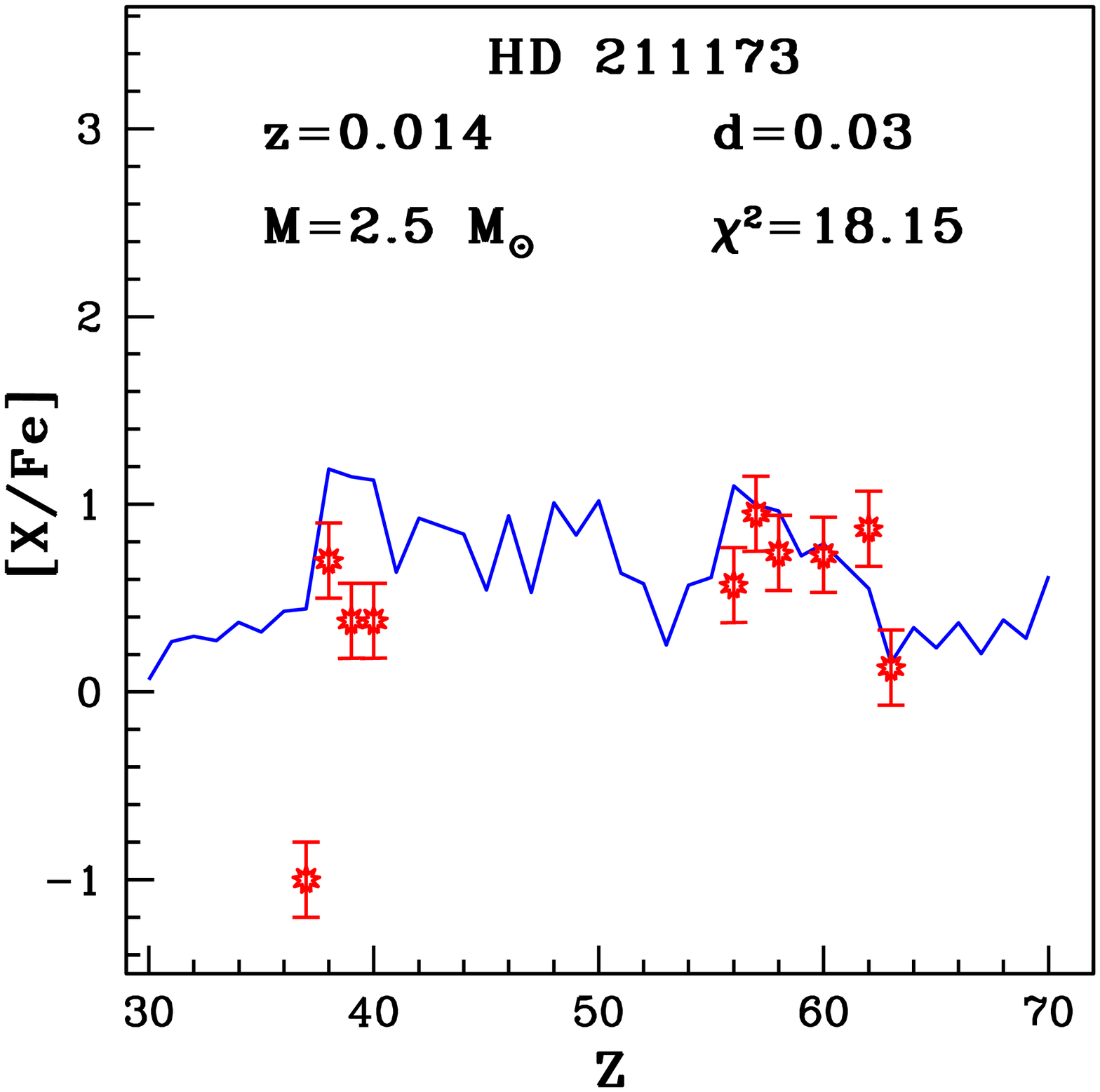}
\caption{Solid curve represent the best fit for the parametric model function.
The points with error bars indicate the observed abundances in the program stars.} 
\label{parametric1}
\end{figure}

We have performed a parametric model based  analysis in order to confirm the mass of the companion AGB stars
by incorporating the dilution experienced by the s-rich material after the mass transfer.
The  dilution factor, d, is defined as M$_{\star}^{env}$/M$_{AGB}^{transf}$ = 10$^{d}$,
where M$_{\star}^{env}$ is the mass of the envelope of the observed star 
after the mass transfer, M$_{AGB}^{transf}$ is the mass transferred from the AGB.
The dilution factor is derived by comparing the observed abundance 
with the predicted abundance from FRUITY model for the heavy elements 
(Rb, Sr, Y, Zr, Ba, La, Ce, Pr, Nd, Sm and Eu). A publicly available 
(http://fruity.oa-teramo.inaf.it/, Web sites of 
the Teramo Observatory (INAF))  data set for the s-process in AGB stars 
is the FRANEC Repository of Updated Isotopic Tables \& Yields (FRUITY) 
models (Cristallo et al. 2009, 2011, 2015b). These models cover the 
whole range  of metallicity observed for Ba stars from z = 0.001 to 
z = 0.020 for the mass range 1.3 - 6.0 M$_{\odot}$. We have compared 
our estimated abundances with the FRUITY model. The solar
values are taken as the initial composition. The observed
elemental abundances are fitted with the parametric model function.
The best fitting masses and corresponding dilution factors along with the
$\chi^{2}$ values are given in Table \ref{parametric model}. 
A few examples of the best fits obtained are shown in Figure \ref{parametric1}.
All the Ba stars are found to have low-mass AGB companions with M $\leq$ 3 M$_{\odot}$.

{\footnotesize
 \begin{table}
\caption{The best fitting mass, dilution factor and reduced chi-square values.}  \label{parametric model}
\resizebox{\columnwidth}{!}{\begin{tabular}{|l|l|l|l|}
\hline                       
star name     & M$_{AGB}$       & d         & $\chi^{2}$  \\
              & (M$_{\odot}$)   &           &                 \\
\hline
HD 32712      & 2.0             & 0.001     & 16.14               \\
HD 36650      & 3.0             & 0.04      & 8.08            \\
HD 179832     & 3.0             & 0.75      & 48.01        \\
HD 211173     & 2.5             & 0.03      & 18.15      \\
\hline
\end{tabular}} 
 \end{table}}

In the FRUITY models,  
a  standard $^{13}$C pocket is being considered, the different 
$^{13}$C pocket efficiencies  may explain the observed discrepancy. 
Also, inclusion of stellar rotations may cause the deviation in the observed abundance pattern
, which is lacking  in the current FRUITY models. Rotation induced mixing is found to alter the 
extend of $^{13}$C pocket, altering the s-process abundance pattern (Langer et al. 1999). 
However, we note that, a study made by 
Cseh et al. (2018) using the rotating star models available for the 
metallicity range of  Ba stars (Piersanti et al. 2013) could not 
reproduce the observed abundance  ratios of stars studied in 
de Castro et al. (2016). 

\subsection{The [hs/ls] ratio}
In addition to the [Rb/Zr] ratio, the [hs/ls] ratio in the Ba stars is an indicator of the neutron source 
and hence the mass of the AGB stars. Here ls refers to the light 
s-process elements (Sr, Y and Zr) and hs to the heavy s-process elements 
(Ba, La, Ce and Nd). The observed [hs/ls] ratio is in the range 0.19 - 1.15, which agrees with the 
model calculations of Busso et al. (2001) for similar metallicities, for low mass AGB stars 
considering $^{13}$C($\alpha$, n)$^{16}$O neutron source. 
This also confirms the lower-mass for the companion AGBs. 

\subsection{Comparison with low-mass AGB abundance}
\begin{figure}
\begin{center}
\includegraphics[scale=0.35]{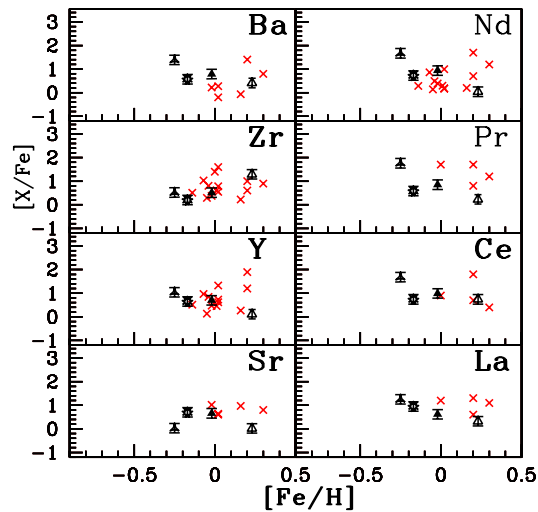}
\caption{\small{Comparison of abundance ratios of neutron-capture elements 
observed in the program stars and the AGB stars with respect to 
metallicity [Fe/H].  Red crosses represent the AGB stars from literature 
(Smith \& Lambert 1985, 1986b, 1990, Abia \& wallerstein 1998).
 HD~32712 (starred triangle), 
HD~36650 (filled triangle),
HD~179832 (open triangle) and
HD~211173 (nine-sided star).}} 
\label{agb_heavy_comparison}
\end{center}
\end{figure}

We have compared the observed  abundance ratios for eight neutron-capture elements 
in the Ba stars with their counterparts in the low-mass AGB stars from literaure, 
that are found to be associated with $^{13}$C($\alpha$, n)$^{16}$O 
neutron source (Figure \ref{agb_heavy_comparison}). The comparison shows a pretty good match 
between the abundances observed in both Ba stars and the AGBs. The  scatter observed in the  ratios may be a 
consequence of different  dilution factors during the mass transfer, 
as well as the orbital  parameters, metallicity and initial mass (de Castro et al. 2016). 

\subsection{Mg abundance in support of neutron source.} 
\begin{figure}
\centering
\includegraphics[scale=0.30]{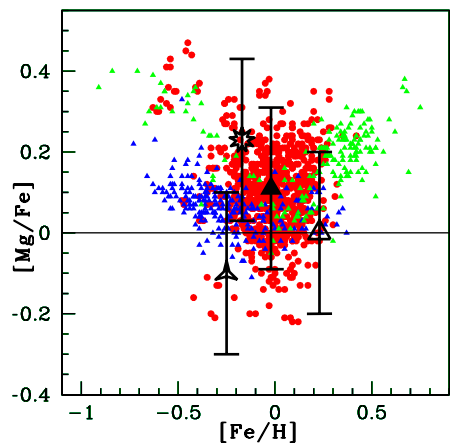}
\caption{Abundances of Mg in our program stars,
HD~32712 (starred triangle), 
HD~36650 (filled triangle),
HD~179832 (open triangle) and
HD~211173 (nine-sided star)
compared with the disk dwarfs and field giants. Red circles represent normal field giants 
from literature (Luck \& Heiter, 2007). 
Blue triangles represent thin disk dwarfs from Bensby et al. (2004), Reddy et al. (2003).
Green triangles represent thick disk dwarfs from Bensby et al. (2004), Reddy et al. (2006).} \label{mg}
\end{figure}

Another check for the companion AGB mass is the Mg abundance. A Mg enrichment is expected to observe in the stars if the s-process over abundance is resulting from the 
neutrons produced during the convective thermal pulses through the reaction $^{22}$Ne($\alpha$,n)$^{25}${Mg}. 
A comparison of the Mg abundances observed in the program stars and that in the 
disk stars and field giants is illustrated in Figure \ref{mg}.
As it is obvious from the figure, we could not find any enhancement of Mg in our sample when compared to the disk stars and normal giants.
This discards the fact that the origin of neutron is $^{22}$Ne($\alpha$,n)$^{25}${Mg} source.

\section{Conclusion}
The detection of Rb I line at 7800.259 {\rm \AA} in the spectra of four program stars allowed us to determine 
        [Rb/Zr] ratio for these objects. The negative values obtained for this ratio in 
        these  stars indicate the operation of $^{13}$C($\alpha$, n)$^{16}$O reaction. As this reaction occurs in the 
        low-mass AGB stars, we confirm that the former companions of these stars are low-mass AGB stars with M $\leq$ 3 M$_{\odot}$. 
 A comparison of observed abundances  with the predictions from  FRUITY models, and with those that are observed  in  low-mass AGB 
        stars from literature, also confirms  low-mass for the former companion AGB stars.   
The observed [hs/ls] ratio  in the range 0.19 - 1.15, that agrees with the
model calculations of Busso et al. (2001)  for  AGB stars
considering $^{13}$C($\alpha$, n)$^{16}$O neutron source, is also an  
indication of  low-mass for the companion AGBs.
   An enhancement of Mg abundances compared to their counterparts in disk stars and normal giants 
         would have indicated the operation of $^{22}$Ne($\alpha$, n)$^{25}$Mg. We did not find any enhancement of Mg  
        in our sample and hence discard the source of neutron as the $^{22}$Ne($\alpha$, n)$^{25}$Mg 
        reaction. This again confirms low-mass for the companion AGBs.

%%Appendix

\section*{Acknowledgements}

Funding from the DST SERB project No. EMR/2016/005283 is gratefully 
acknowledged. 
This work made use of the SIMBAD astronomical database, operated
at CDS, Strasbourg, France, and the NASA ADS, USA.
This work has made use of data from the European Space Agency (ESA) 
mission Gaia (https://www.cosmos.esa.int/gaia), processed by the Gaia 
Data Processing and Analysis Consortium 
(DPAC, https://www.cosmos.esa.int/web/gaia/dpac/consortium).
\vspace{-1em}

%%use \balance somewhere in the left column of the last page to balance the two columns in the end page

%%References section

\end{document}